\pdfoutput=1
\documentclass[pdflatex,sn-mathphys]{sn-jnl}% Math and Physical Sciences Reference Style
\theoremstyle{thmstyleone}%
\theoremstyle{thmstyletwo}%

\theoremstyle{thmstylethree}%
\usepackage[numbers]{natbib}

\begin{document}

\title[Towards overcoming data scarcity with mixture of experts]{Towards overcoming data scarcity in materials science: unifying models and datasets with a mixture of experts framework}


\author*[1]{\fnm{Rees} \sur{Chang}}\email{reeswc2@illinois.edu}

\author*[2]{\fnm{Yu-Xiong} \sur{Wang}}\email{yxw@illinois.edu}

\author*[3,4]{\fnm{Elif} \sur{Ertekin}}\email{ertekin@illinois.edu}

\affil[1]{\orgdiv{Department of Materials Science and Engineering}, \orgname{University of Illinois at Urbana-Champaign}, \orgaddress{\street{1304 West Green Street}, \city{Urbana}, \postcode{61801}, \state{Illinois}, \country{United States of America}}}

\affil[2]{\orgdiv{Department of Computer Science}, \orgname{University of Illinois at Urbana-Champaign}, \orgaddress{\street{201 North Goodwin Avenue}, \city{Urbana}, \postcode{61801}, \state{IL}, \country{United States of America}}}

\affil[3]{\orgdiv{Materials Research Laboratory}, \orgname{University of Illinois at Urbana-Champaign}, \orgaddress{\street{104 South Goodwin Avenue}, \city{Urbana}, \postcode{61801}, \state{IL}, \country{United States of America}}}

\affil[4]{\orgdiv{Department of Mechanical Science and Engineering}, \orgname{University of Illinois at Urbana-Champaign}, \orgaddress{\street{1206 West Green Street}, \city{Urbana}, \postcode{61801}, \state{IL}, \country{United States of America}}}


\abstract{While machine learning has emerged in recent years as a useful tool for rapid prediction of materials properties, generating sufficient data to reliably train models without overfitting is still impractical for many applications. Towards overcoming this limitation, we present a general framework for leveraging complementary information across different models and datasets for accurate prediction of data scarce materials properties. Our approach, based on a machine learning paradigm called mixture of experts, outperforms pairwise transfer learning on 16 of 19 materials property regression tasks, performing comparably on the remaining three. Unlike pairwise transfer learning, our framework automatically learns to combine information from multiple source tasks in a single training run, alleviating the need for brute-force experiments to determine which source task to transfer from. The approach also provides an interpretable, model-agnostic, and scalable mechanism to transfer information from an arbitrary number of models and datasets to any downstream property prediction task. We anticipate the performance of our framework will further improve as better model architectures, new pre-training tasks, and larger materials datasets are developed by the community.}




\maketitle

\section{Introduction}\label{sec1}

In recent years, software development of automated density functional theory (DFT) calculation workflows has led to the emergence of large open-source databases of materials and their simulated properties  \cite{materialsproject2013,aflow2012,jarvis2020}. However, due to computational restraints, not all properties are computed for all materials in these databases. For example, at the time of writing, the Materials Project (MP) \cite{materialsproject2013} contains 144,595 inorganic materials, but only 76,240 electronic bandstructures, 14,072 elastic tensors, and 3,402 piezoelectric tensors. Many studies have thus trained supervised machine learning (ML) models on materials for which property data is available, subsequently screening the remaining materials orders of magnitude faster than DFT. After identifying promising materials with ML-based screenings, identified materials are studied more rigorously with DFT and/or experiment. Example applications wherein ML-based screenings led to successful simulated or experimental validation include photovoltaics, superhard materials, batteries, hydrogen storage materials, ferroelectrics, shape memory alloys, dielectrics, and more \cite{successes_ml_for_materials}.

To handle the data types encountered in materials, ML approaches in materials science generally involve statistical learning models using hand-crafted, application-dependent descriptors as input \cite{materials_logistic_regression, matbench2020} or graph neural networks (GNNs) directly using materials' atomic structures as input \cite{megnet2019,cgcnn2018}. While the latter models have shown superior performance likely by more faithful representation of atomic structures \cite{critical_examination2020}, their large number of trainable parameters requires on the order of $10^4$ data examples to sufficiently reduce overfitting relative to descriptor-based methods \cite{matbench2020}. Acquiring $10^4$ data examples can be impractically expensive, limiting our ability to build predictive ML models, e.g., for experimental data and complex systems like layered materials, surfaces, and materials with point defects. Similarly, generating large amounts of data is infeasible for rare materials behaviors and phases like high temperature superconductors or spin liquids. Developing predictive ML models to effectively handle data scarcity in materials science is thus a pervasive challenge with practical significance for a range of technologies.

Several approaches have been applied in materials science to reduce the large data requirement of neural networks. Many of these approaches can be classified as regularizing neural networks to perform well across \emph{multiple} relevant tasks - similar to how humans use background knowledge to learn from few examples. 

One such regularization technique commonly employed in materials science is pairwise transfer learning (TL), wherein parameters of a model pre-trained on a data-abundant source task (e.g., predicting formation energy) are used to initialize training on a data-scarce, downstream task (e.g., predicting experimental band gaps) \cite{crysxpp2022, atomsets2021, multifidelity_bandgap2_2021, multifidelity_bandgap2021, multifidelity_eform2019, megnet2019, tl_cgcnn2021, materials_transfer2017}.
To avoid \emph{catastrophic forgetting} of the source task during transfer to the downstream task, early layers of the pre-trained model are typically frozen while later layers are fine-tuned, i.e., updated with a reduced learning rate \cite{tl_cgcnn2021}. However, TL suffers from several limitations; its success is contingent on the existence of a source task with many data examples and high similarity to the downstream task. Additionally, TL only allows information to be leveraged from a single task, and the source task from which to transfer from is not generally known \emph{a priori} \cite{taskonomy2018, nlp_transferability2020}. Previous studies have thus either transferred from the largest available source task \cite{megnet2019}, transferred from lower to higher fidelity data of the same property \cite{multifidelity_bandgap2_2021, multifidelity_bandgap2021, multifidelity_eform2019}, conducted brute-force experiments on different source tasks \cite{atomsets2021, shotgun_transfer2019}, or engineered new source tasks not directly relevant for a materials application but serving as generalizable pre-training tasks \cite{crysxpp2022, contrastive_ssl_materials2022}.

Other techniques, such as multitask learning (MTL), can leverage information across many tasks. MTL has already been used to improve model performance by jointly predicting formation energies, band gaps, and Fermi energies with a single model \cite{multitask_cgcnn_2018}. However, MTL models are in general difficult to train; determining task groupings for joint training without detriment to performance (i.e., without \emph{negative transfer} from \emph{task interference}) is an open research question \cite{crosstask_consistency2020, taskgroupings2020}. Furthermore, optimal groupings are sensitive to hyperparameters like learning rate and batch size \cite{finn_task_groupings}. This sensitivity arises because MTL models must overcome imbalanced task gradient magnitudes and conflicting task gradient directions during training \cite{gradDrop2020, gradnorm2018, rotograd2022}. Also, MTL models frequently suffer from catastrophic forgetting when adapted to new tasks \cite{multitask_catastrophic_forgetting}. 

In this work, to overcome the aforementioned limitations of TL and difficulties of MTL, we propose a mixture of experts (MoE) framework for materials property prediction.  By construction, our framework can leverage information from an arbitrary number of source tasks and pre-trained models to any downstream task, does not experience catastrophic forgetting or task interference across source tasks as in MTL, and automatically learns which source tasks and pre-trained models are the most useful for a downstream task in a single training run. Our framework consistently outperforms pairwise TL on a suite of data scarce property prediction tasks; emits interpretable relationships between source and downstream property prediction tasks; and provides a general, modular framework to combine complementary models and datasets for data scarce property prediction. The generality of our approach also makes it compatible with any new source tasks, model architectures, or datasets which may be developed in the future.

\section{Results and Discussion}\label{sec2}

\subsection{Pairwise Transfer Learning}
Pairwise transfer learning involves using all or a subset of parameters from a pre-trained model to initialize training on a data-scarce, downstream task. We know fundamental rules of quantum chemistry generalize across materials and properties, i.e., the periodic table and Schrodinger's equation are universal. 
However, the final mapping from fundamental physics to a specific property depends heavily on the property. For example, while both formation energy and electronic band gap can be obtained from DFT, computing formation energy requires comparing to a relevant reference state, while computing band gaps requires comparing band edge positions. 
Thus, after pre-training a model on a source task for TL (and MoE), we only re-used a subset of the pre-trained model parameters to produce generalizable features of an atomic structure. We let these pre-trained parameters define a feature \emph{extractor}, $E(\cdot)$. Specifically, the extractor takes in an atomic structure $x$ and outputs a feature vector $E(x)$ describing the structure. Predictions of a scalar property of any atomic structure is then produced by passing the feature vector $E(x)$ through a property-specific \emph{head} neural network, $H(\cdot)$. Putting it together, predictions $\hat{y}$ are produced as $\hat{y} = H(E(x))$.

Similar to Sanyal et al. \cite{multitask_cgcnn_2018} and Xie et al. \cite{tianxie_experimental_tl}, we let the extractor $E(\cdot)$ be the atom embedding and graph convolutional layers of a crystal graph convolutional neural network (CGCNN) \cite{cgcnn2018}. These layers produce a representation of a crystal from its constituent atom types and pairwise interatomic distances. Our head $H(\cdot)$ is a multilayer perceptron. Specific hyperparameters of the architecture can be found in Table \ref{model_hyperparameters}. In our pairwise TL experiments, we found it beneficial to extract from and fine-tune the last graph convolutional layer when transfer learning to a downstream task (see Figures \ref{layer_to_extract_from} and \ref{n_layers_to_finetune}). We applied these design choices to all TL and MoE experiments in the rest of this paper.

\subsection{The Mixture of Experts framework}\label{subsec2}
    MoEs were first introduced more than three decades ago \cite{original_moe1991} and have since been studied as a general-purpose neural network layer notably for tasks in natural language processing \cite{shazeer_moe2017}. MoE layers consist of multiple \emph{expert} neural networks and a trainable gating network which, often conditionally, routes inputs through the experts. The output of the MoE layer is then computed by aggregating outputs of all the activated experts. A result of the MoE layer's gating mechanism is that large parts of the model can be inactive on a per-example basis, enabling massive increases in model capacity and performance without concomitant increases in training cost. Interestingly, in natural language processing, it has also been shown that the experts tend to automatically become highly specialized based on syntax and semantics \cite{shazeer_moe2017}.
    
    Formally, a MoE layer consists of $m$ experts $E_{\phi_1},...,E_{\phi_m}$ parameterized by $\phi_1,...,\phi_m$ and a gating function $G(x, \theta, k)$ which takes in trainable parameters $\theta$ and produces a $k$-sparse, $m$-dimensional probability vector. In our work, since each expert is responsible for producing a feature vector describing a material, we refer to each expert as an \emph{extractor}. For simplicity, we also chose to make our gating function independent of the model input (i.e., which material we are making a property prediction for), so we have $G(x,\theta,k) = G(\theta,k)$. For a given input $x$, we denote the output of the $i$-th extractor as $E_{\phi_i}(x)$ and the $i$-th output of $G(\theta, k)$ as $G_i(\theta, k)$. The output $f$ of our MoE layer is a feature vector produced by a mixture of extractors, i.e.,
    
    \begin{equation}\label{moe_output_eqn}
        f = \bigoplus_{i=1}^m G_i(\theta, k) E_{\phi_i}(x) \hspace{0.5em},
    \end{equation}
    
\noindent where $\bigoplus$ is an aggregation function. We experimented with letting the aggregation function be concatenation or addition, comparing performance with end-to-end learning of a weighted ensemble of different fine-tuned CGCNN predictions. Table \ref{mixing_methods_table} reports the mean absolute error (MAE) of each method on three data scarce tasks: predicting piezoelectric moduli \cite{piezo2015}, 2D exfoliation energies \cite{jarvis_2d_2017}, and experimental formation energies \cite{expt_formation_energy_kingsbury, kim_expt_ef2017}. These tasks consisted of 941, 636, and 1,709 data examples, respectively. None of the three aggregation methods consistently outperformed the others on these tasks, so we opted for addition. An advantage of this choice is that the model's feature dimensionality becomes independent of the number of feature extractors. As a proof-of-concept, our extractors $\{E_{\phi_i}(\cdot)\}$ are CGCNNs each pre-trained on a different materials property dataset with at least $10^4$ examples. All datasets were acquired through \texttt{Matminer} \cite{matminer2018}.
    
    \begin{table}[h]
    \begin{center}
    \begin{minipage}{230pt}
    \caption{Average test mean absolute error (MAE) over 5 random seeds for different methods of aggregating pre-trained extractors. Smaller MAEs are better. Each method is benchmarked on predicting piezoelectric modulus ($d_{33}$) \cite{piezo2015}, 2D exfoliation energies ($E_\mathrm{exfol}$) \cite{jarvis_2d_2017}, and experimental formation energies (Expt. $E_f$) \cite{kim_expt_ef2017, expt_formation_energy_kingsbury}.}\label{mixing_methods_table}%
    \begin{tabular}{@{}llll@{}}
    \toprule
      & $d_{33}$ & $E_\mathrm{exfol}$ & Expt. $E_f$\\
      &          &  (meV/at)          & (eV/at)\\
    \midrule
    Ensemble     & $0.222 \pm 0.027$   & $51.6 \pm 10.4$ & $0.118 \pm 0.011$\\
    Concatenate  & $0.262 \pm 0.049$ & $56.3 \pm 7.4$ & $0.0870 \pm 0.0118$\\
    Add          & $0.227 \pm 0.040$  & $52.6 \pm 12.7$  & $0.0936 \pm 0.0096$\\
    \botrule
    \end{tabular}
    \end{minipage}
    \end{center}
    \end{table}
    
    We parameterize our gating mechanism with $\theta \in \mathbb{R}^m$ and a hyperparameter controlling sparsity, $k \in \mathbb{N}^+$, where $\mathbb{N}^+$ denotes natural numbers greater than zero. Our gating function $G(\theta, k)$ is as follows:
    \begin{align}
        G(\theta, k) &= \mathrm{Softmax}(\mathrm{KeepTopK}(\theta, k)) \hspace{0.5em}, \label{eqG}\\
        \mathrm{KeepTopK}(\theta, k)_i &= \left\{ 
                      \begin{array}{ c l }
                        \theta_i    & \quad \textrm{if } \theta_i \textrm{ is in the top } k \textrm{ elements of } \theta \\
                        -\infty     & \quad \textrm{otherwise.}
                      \end{array}
                    \right. \hspace{0.5em} \label{eqTopK}
    \end{align}
    
    \noindent Before applying the Softmax function, we only keep the top $k$ values of $\theta$. Mathematically, this is equivalent to setting the rest of the values to $-\infty$, assigning the corresponding extractors a gating value of 0 after applying the Softmax. To encourage the model to focus on the most relevant extractors, we followed the parameter-free method of Lin et al.\ \cite{attention_score_regularizer} and added a regularization term $P$ to the training loss:
    \begin{align}
        P = \lambda(\boldsymbol{a}^T \boldsymbol{a} - 1)^2 \hspace{0.5em}. \label{regularization_eq}
    \end{align}
    Here, $\boldsymbol{a}=G(\theta,k)\in \mathbb{R}^{(m \times 1)}$ is a vector of probability scores assigned to each extractor, and $\lambda$ is a hyperparameter weighting the regularization term. We set $\lambda=0.01$. Intuitively, $(\boldsymbol{a}^T\boldsymbol{a}-1)^2 \geq 0$ with equality if and only if $\boldsymbol{a}$ concentrates all probability mass on a single extractor (wherein $\boldsymbol{a}^T\boldsymbol{a}=1$).
    
    While we chose the extractors $E_{\phi_1},...,E_{\phi_m}$ to be CGCNNs, this choice can be much more flexible. For example, extractor outputs could be embeddings from other graph neural networks \cite{megnet2019, cgcnn2018, e3nn_phonon_dos2020}, hand-crafted featurizers \cite{matbench2020}, language models \cite{materials_word_embeddings2019}, or generative models \cite{cdvae2022} trained by single-task, multitask, supervised, unsupervised, semi-supervised, or self-supervised learning. This ability to combine different extractor architectures is distinct from TL or MTL, where single architectures must be used across all tasks. Additionally, extractors can be trained on any dataset which serves as generalizable pre-training data. These might include materials properties from the Materials Project \cite{materialsproject2013}, the Open Quantum Materials Database \cite{oqmd2013}, JARVIS \cite{jarvis2020}, or AFLOW \cite{aflow2012}; text from scientific journal abstracts \cite{materials_word_embeddings2019}; or data generated for unsupervised or self-supervised learning \cite{cdvae2022, crysxpp2022, contrastive_ssl_materials2022}.
    
\begin{figure}[h]%
    \centering
    \includegraphics[width=1.0\textwidth]{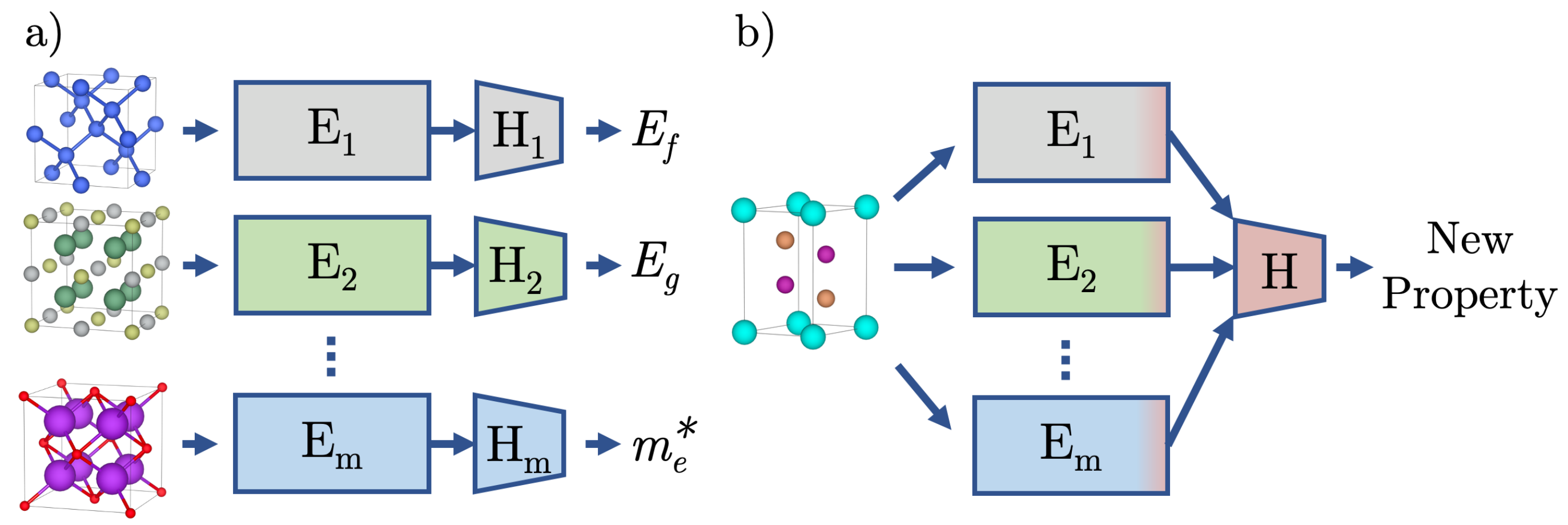}
    \caption{a) Schematic of pre-training the feature extractors for the mixture of experts framework. Separate machine learning models, each consisting of a feature extractor and a classification or regression head, are trained on separate learning tasks. b) Schematic of the pre-trained experts being adapted to a downstream learning task along with a newly initialized head.}\label{moe_schematic}
    \end{figure}

    A schematic summarizing our framework can be seen in Fig. \ref{moe_schematic}. The first phase of our approach is to pre-train separate models on different \emph{source tasks}. In our case, these source tasks involve prediction of scalar properties from large datasets, where, following Ref. \cite{matbench2020}, we defined ``large'' as consisting of more than $10^4$ examples (Fig. \ref{moe_schematic}a). A complete list of our source tasks and the resulting pre-trained model performances are enumerated in the supplementary information (Table \ref{extractor_tasks}). Because each extractor is pre-trained separately, there is no possibility of task interference during pre-training as in MTL. The second phase is to combine and adapt the separate models to a downstream, data scarce task (Fig. \ref{moe_schematic}b). Our adaptation process consisted of training a randomly initialized head while fine-tuning the last layer of each extractor towards the new task.

    \subsubsection{MoE outperforms TL from the best pre-trained model}\label{subsubsec2}

    We examined whether transferring information from multiple pre-trained models with the MoE framework could outperform transferring from a single pre-trained model.  Specifically, we compared single-task learning (STL) from a randomly initialized model, TL from the best of three expertly chosen source tasks (Best TL-(3)), MoE using all three expertly chosen source tasks (MoE-(3)), and MoE using all available source tasks (MoE-(18)). 
    
    As before, each method was evaluated on three downstream, data scarce tasks: predicting piezoelectric modulus \cite{piezo2015}, 2D materials' exfoliation energies \cite{jarvis_2d_2017}, and experimental formation energies \cite{kim_expt_ef2017, expt_formation_energy_kingsbury}. For TL and MoE, we let MP formation energies be a pre-training task across all three downstream tasks since it is the largest dataset available in \texttt{Matminer}. For Best TL-(3) and MoE-(3), the remaining pre-training tasks were chosen as follows. For predicting piezoelectric modulus, we transferred from extractors pre-trained on predicting MP bulk moduli \cite{mp_elastic2015, materialsproject2013} since piezoelectric tensors are derived in part from elastic tensors \cite{piezo2015}, and MP band gaps \cite{materialsproject2013} since a nonzero band gap is required to maintain an electric polarization. For predicting 2D materials' exfoliation energies, we transferred from JARVIS formation energies since these are also thermodynamic quantities and come from the same data source \cite{jarvis2020}, and JARVIS shear moduli since small elastic constants have been suggested as a signature of weak van der Waals bonding and low exfoliation energies \cite{jarvis_3d_dft2018}. For predicting experimental formation energies, we transferred from JARVIS \cite{jarvis_3d_dft2018, jarvis2020} and MP perovskite formation energies \cite{castelli2012}.
    
    Unsurprisingly, Best TL-(3) outperformed STL on all three downstream tasks. Notably for each downstream task, MoE-(3) performs as well or better than transferring from the best individual pre-trained model (Best TL-(3) in Table \ref{learn_to_weight_extractors}). This result is despite the former approach fine-tuning three times the number of pre-trained parameters, wherein one might expect overfitting on data scarce tasks.
    
    In general, relationships between source and downstream tasks may be counterintuitive or unknown. For example, we may lack or have incorrect scientific understanding for a particular property. Or, if source tasks are unsupervised or self-supervised, relationships to properties may not be justifiable with domain knowledge. To handle these scenarios, we separately pre-trained eighteen extractors on different source tasks, challenging a MoE model to automatically learn which extractors were most useful for the downstream task at hand (MoE-(18) in Table \ref{learn_to_weight_extractors}). We observed improved performance over using three expertly chosen extractors with Best TL-(3) and MoE-(3) for predicting piezoelectric modulus and experimental formation energies, as well as comparable performance when predicting exfoliation energy. A full list of the source tasks is in the supplementary information.
    
    \begin{table}[h]
    \begin{center}
    \begin{minipage}{250pt}
    \caption{Average test MAE over 5 random seeds for single-task learning, pairwise transfer learning, and mixture of experts using expertly chosen and all available pre-training tasks. Best MAEs are bolded. See Table \ref{maes_main_table} for data source references.}\label{learn_to_weight_extractors}%
    \begin{tabular}{@{}llll@{}}
    \toprule
        & $d_{33}$ & $E_\mathrm{exfol}$ & Expt. $E_f$\\
        &                     & (meV/at)                    & (eV/at)\\
    \midrule
    STL    & $0.228\pm0.033$   & $62.0\pm13.6$  & $0.114\pm0.006$\\
    Best TL-(3) & $0.234\pm0.021$\footnotemark[1]   & $53.0\pm10.6$\footnotemark[2]  & $0.102\pm0.015$\footnotemark[3]\\
    MoE-(3)    & $0.227\pm0.040$   & $\boldsymbol{52.6\pm10.4}$  & $0.0936\pm0.0096$\\
    MoE-(18)  & $\boldsymbol{0.208\pm0.029}$ & $54.4\pm10.7$ & $\boldsymbol{0.0895\pm0.0099}$ \\
    \botrule
    \end{tabular}
    \footnotetext[1]{TL from MP bulk moduli outperformed TL from MP formation energies and MP band gaps with MAEs of $0.294\pm0.097$ and $0.239\pm0.030$, respectively.}
    \footnotetext[2]{TL from JARVIS formation energies outperformed TL from MP formation energies and shear moduli with MAEs of $60.4\pm9.5$ and $66.1\pm7.7$, respectively.}
    \footnotetext[3]{TL from MP formation energies outperformed TL from JARVIS and perovskite formation energies with MAEs of $0.108\pm0.013$ and $0.180\pm0.010$, respectively.}
    \end{minipage}
    \end{center}
    \end{table}
    
    \subsubsection{Benchmarking MoE}
    To evaluate our MoE framework's ability to handle data scarcity, we assessed the framework on 19 materials property regression tasks from \texttt{Matminer} \cite{matminer2018} with dataset sizes ranging from 120 to 8,043 examples. The task datasets span thermodynamic, electronic, mechanical, and dielectric properties; intrinsic and extrinsic properties at fixed temperature and doping concentration; as well as data generated with various DFT exchange-correlation functionals and physical experiments. We set the sparsity hyperparameter $k$ in Eq. \eqref{eqTopK} to 18, allowing the model to utilize all source tasks. Parameter vector $\theta$ (Eq. \eqref{eqG}) was initialized to a vector of ones, corresponding to a uniform distribution of probability scores assigned to each pre-trained model. Hyperparameters were held fixed across all tasks.
    
    \begin{table}[h]
    \begin{center}
    \begin{minipage}{315pt}
    \caption{Average test MAE over 5 random seeds on 19 data scarce regression tasks for single-task learning, transfer learning, and mixture of experts.}\label{maes_main_table}%
    \begin{tabular}{@{}llll@{}}
    \toprule
    Downstream Task                             & STL                   & TL from MP $E_f$      & MoE (ours)\\
    (Dataset size)\\
    \midrule
    Expt. $E_f$\footnotemark[1] (1,709)                         & $0.114\pm0.006$       & $0.102\pm0.015$       & $\boldsymbol{0.0908\pm0.0142}$ \\
    $E_\mathrm{exfol}$\footnotemark[2] (636)                    & $62.0\pm13.6$         & $60.4\pm9.5$          & $\boldsymbol{53.6\pm10.5}$ \\
    $d_{33}$\footnotemark[3] (941)                              & $0.228\pm0.033$       & $0.294\pm0.097$       & $\boldsymbol{0.208\pm0.029}$ \\
    PhonDOS peak\footnotemark[4] (1,265)                        & $0.126\pm0.014$       & $\boldsymbol{0.110\pm0.010}$       & $0.115\pm0.017$\\
    2D $E_f$\footnotemark[5] (633)                              & $0.165\pm0.024$       & $0.146\pm0.021$       & $\boldsymbol{0.145\pm0.011}$ \\
    2D $E_g$, Opt\footnotemark[6] (522)                         & $0.693\pm0.148$       & $0.698\pm0.198$       & $\boldsymbol{0.530\pm0.101}$ \\
    2D $E_g$, Tbmbj\footnotemark[7] (120)                       & $1.31\pm0.29$         & $1.77\pm0.38$         & $\boldsymbol{1.19\pm0.26}$ \\
    $A^U$\footnotemark[8] (1,181)                            & $3.69\pm2.82$         & $3.16\pm2.05$         & $\boldsymbol{3.06\pm2.31}$ \\
    Log($\epsilon_\infty$)\footnotemark[9] (1,296)                & $0.170\pm0.039$       & $0.163\pm0.032$       & $\boldsymbol{0.161\pm0.035}$ \\
    Log($\epsilon_\mathrm{total}$)\footnotemark[10] (1,296)        & $0.254\pm0.038$        & $\boldsymbol{0.239\pm0.030}$       & $0.244\pm0.027$\\
    Poisson Ratio\footnotemark[11] (1,181)                       & $0.0325\pm0.0003$     & $0.0314\pm0.0015$     & $\boldsymbol{0.0293\pm0.0020}$ \\
    $\epsilon_\mathrm{poly}^\infty$\footnotemark[12] (1,056)  & $2.94\pm0.89$         & $3.05\pm0.72$         & $\boldsymbol{2.58\pm0.84}$ \\
    $\epsilon_\mathrm{poly}$\footnotemark[13] (1,056)         & $6.40\pm1.54$         & $6.58\pm0.95$         & $\boldsymbol{5.60\pm1.51}$ \\
    2D n, Opt\footnotemark[14] (522)                             & $3.66\pm0.44$         & $3.76\pm0.30$         & $\boldsymbol{2.99\pm0.53}$ \\
    2D n, Tbmbj\footnotemark[15] (120)                           & $\boldsymbol{10.7\pm20.6}$         & $34.2\pm41.6$         & $40.8\pm33.6$\\
    3D n, PBE\footnotemark[16] (4,764)                             & $0.0860\pm0.0131$     & $0.0869\pm0.0106$     & $\boldsymbol{0.0823\pm0.0108}$ \\
    Expt. $E_g$\footnotemark[17] (2,481)                         & $0.460\pm0.046$       & $0.444\pm0.030$       & $\boldsymbol{0.371\pm0.042}$ \\
    $\epsilon_\mathrm{avg}$, Tbmbj\footnotemark[18] (8.043)      & $32.7\pm2.6$          & $47.1\pm6.7$          & $\boldsymbol{27.3\pm3.7}$ \\
    $E_g$, Tbmbj\footnotemark[19] (7,348)                        & $0.503\pm0.018$       & $0.462\pm0.010$       & $\boldsymbol{0.385\pm0.013}$ \\
    \botrule
    \end{tabular}
    \footnotetext[1]{Experimental formation enthalpies (eV/atom) from \texttt{Matminer}'s \texttt{expt\_formation\_enthalpy} \cite{kim_expt_ef2017} and \texttt{expt\_formation\_enthalpy\_kingsbury} datasets \cite{expt_formation_energy_kingsbury}. The former was preferred when duplicates arose.}
    \footnotetext[2]{Exfoliation energies (meV/atom) from \texttt{Matminer}'s \texttt{jarvis\_dft\_2d} dataset \cite{jarvis_2d_2017}.}
    \footnotetext[3]{Piezoelectric modulus from \texttt{Matminer}'s \texttt{piezoelectric\_tensor} dataset \cite{piezo2015}.}
    \footnotetext[4]{Highest frequency optical phonon mode peak (cm$^{-1}$) from \texttt{Matminer}'s \texttt{matbench\_phonons} dataset \cite{phonon_dielectric2018}.}
    \footnotetext[5]{Formation energies (eV/atom) from \texttt{Matminer}'s \texttt{jarvis\_dft\_2d} dataset \cite{jarvis_2d_2017}.}
    \footnotetext[6]{Band gap of 2D materials (eV) from \texttt{Matminer}'s \texttt{jarvis\_dft\_2d} dataset \cite{jarvis_2d_2017}, calculated with the OptB88vDW DFT functional.}
    \footnotetext[7]{Band gap of 2D materials (eV) from \texttt{Matminer}'s \texttt{jarvis\_dft\_2d} dataset \cite{jarvis_2d_2017}, calculated with the TBMBJ DFT functional.}
   \footnotetext[8]{Elastic anisotropy index from \texttt{Matminer}'s \texttt{elastic\_tensor\_2015} dataset \cite{materialsproject2013}.}
    \footnotetext[9]{Electronic contribution to dielectric constant from \texttt{Matminer}'s \texttt{phonon\_dielectric\_mp} dataset\cite{phonon_dielectric2018}.}
    \footnotetext[10]{Dielectric constant from \texttt{Matminer}'s \texttt{phonon\_dielectric\_mp} dataset \cite{phonon_dielectric2018}.}
    \footnotetext[11]{Poisson ratio from \texttt{Matminer}'s \texttt{elastic\_tensor\_2015} dataset \cite{mp_elastic2015}.}
    \footnotetext[12]{Average eigenvalue of the dielectric tensor's electronic component from \texttt{Matminer}'s \texttt{dielectric\_constant} dataset \cite{poly_dielectric2017}.}
    \footnotetext[13]{Average dielectric tensor eigenvalue from \texttt{Matminer}'s \texttt{dielectric\_constant} dataset \cite{poly_dielectric2017}.}
    \footnotetext[14]{Dielectric constant of 2D materials, computed by the OptB88vDW functional, from \texttt{Matminer}'s \texttt{jarvis\_dft\_2d} dataset \cite{jarvis_2d_2017}.}
    \footnotetext[15]{Dielectric constant of 2D materials, computed by the TBMBJ functional, from \texttt{Matminer}'s \texttt{jarvis\_dft\_2d} dataset \cite{jarvis_2d_2017}.}
    \footnotetext[16]{Refractive index from \texttt{Matminer}'s \texttt{dielectric\_constant} dataset \cite{matbench2020,poly_dielectric2017}.}
    \footnotetext[17]{Experimental band gaps (eV) from \texttt{Matminer}'s \texttt{expt\_gap\_kingsbury} dataset \cite{matminer2018}.}
    \footnotetext[18]{Average eigenvalue of dielectric tensor, calculated with the TBMBJ DFT functional, from the \texttt{jarvis\_dft\_3d} database in \texttt{Matminer} \cite{jarvis_3d_dft2018}.}
    \footnotetext[19]{Band gap (eV), calculated with the TBMBJ DFT functional, from \texttt{Matminer}'s \texttt{jarvis\_dft\_3d} dataset \cite{jarvis_3d_dft2018}.}
    \end{minipage}
    \end{center}
    \end{table}
    \clearpage
    
    Mean absolute errors for MoE, single-task learning without pre-training (STL), and TL are reported in Table \ref{maes_main_table}. To avoid expensive, brute-force trial and error of every source task for every downstream task during TL, we used the common heuristic of transferring from the largest available source task, Materials Project (MP) formation energies \cite{materialsproject2013}. This TL strategy has been employed by several works suggesting TL performance for property prediction improves when the model is pre-trained with more data \cite{megnet2019, atomsets2021, tl_cgcnn2021}.
    
    Across the 19 tasks, MoE achieved state-of-the-art performance in 16 of 19 target properties and comparable results on the remaining three properties. Notably, TL performed worse than single-task training from random initialization (STL) on 9 of the 19 tasks. These include predicting piezoelectric modulus, 2D materials' band gaps, and other tasks related to dielectric properties. A possible explanation is that representations trained on formation energies of 3D bulk crystals do not transfer well to different domains (e.g., 2D materials) or dissimilar properties (e.g., piezoelectric or dielectric). In contrast, MoE achieved the best performance on 8 of 9 tasks that pairwise TL experienced negative transfer on, as well as comparable performance with STL on the remaining task; this highlights MoE's ability to avoid negative transfer without the need for any task-specific hyperparameter tuning.
    
    \begin{figure}[h]%
    \centering
    \includegraphics{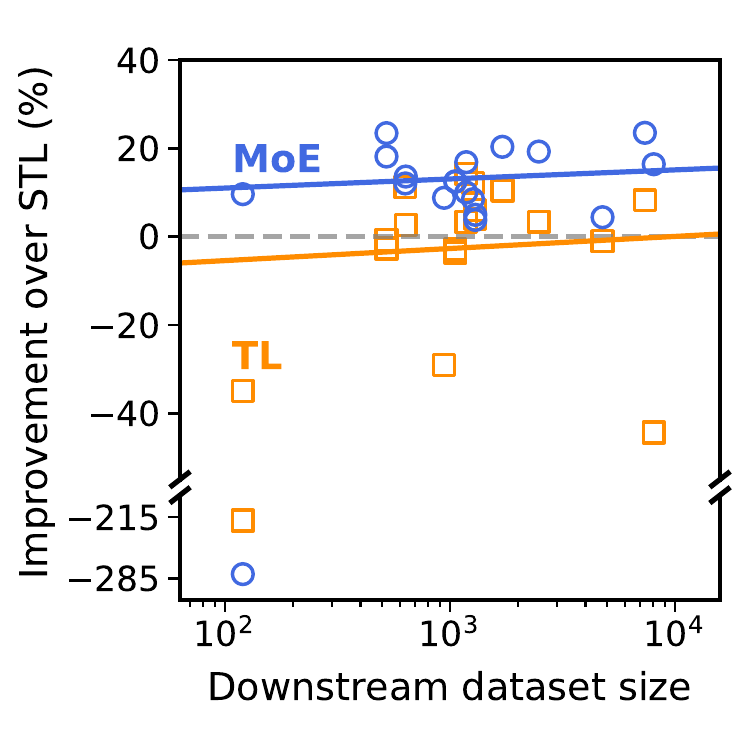}
    \caption{Percent improvement on MAE across 19 materials property regression tasks of transfer learning from Materials Project formation energies (TL) and our mixture of experts approach (MoE) over single-task learning with random initialization (STL). We plot linear-log best fit lines ignoring performance on 2D materials' dielectric constants computed with the TBMBJ functional \cite{jarvis_2d_2017}. For clarity, the y-axis above and below the break have different scales. Examples of positive and negative transfer are indicated by points above and below the gray dotted line, respectively.}\label{improvement_over_STL}
    \end{figure}
    
    For one task, predicting 2D dielectric constants computed by the TBMBJ functional (\emph{2D n, Tbmbj}), MoE (and TL) exhibited worse average performance than STL. Plausible factors underlying the poor performance are the task's dataset size of only 120 examples; the dataset's composition of 2D materials, in contrast to 3D crystals which comprised all the source task datasets; and that the predicted quantity, dielectric constants, varies over a range of hundreds to thousands (as opposed to, e.g., band gaps, which only varies up to tens of eVs). We quantitatively explore the effects of dataset size, domain, and label distribution shifts on MoE performance in Section \ref{understanding_transfer}, where we indeed found that in addition to the smallest dataset size, \emph{2D n, Tbmbj} has both the largest domain and label shifts from the source tasks. Despite these factors, the average MAE of STL on \emph{2D n, Tbmbj} still lies within a standard deviation of the MoE's average MAE. Possible avenues for future improvement include task-specific hyperparameter tuning, inclusion of more generalizable source tasks, and/or development of ML methods which transfer information from more relevant datasets (e.g., other 2D materials properties or dielectric properties). 
    
    Ignoring the \emph{2D n, Tbmbj} dataset, Figure \ref{improvement_over_STL} depicts linear-log best fit lines of TL and MoE improvement over MAEs over STL as a function of downstream dataset size. While the best fit line for TL largely lies below an improvement percent of zero, signifying a tendency for negative transfer, the best fit line for MoE lies entirely above zero. This result highlights MoE's ability to yield positive transfer over the entire range of downstream dataset sizes.
    
    \subsubsection{Understanding positive and negative transfer} \label{understanding_transfer}
    Negative transfer is a pervasive phenomenon in ML wherein transferring information from a source task(s) to a downstream task exhibits worse performance than training on the downstream task from random initialization. Wang et al. \cite{understanding_negative_transfer} and Gong et al. \cite{understanding_transfer2} discussed the key factors from which negative transfer arises: divergence between the source and downstream tasks' joint distributions over the domain and label spaces as well as the size of the labeled downstream task data. Formally, we denote $P_S(X,Y)$ and $P_T(X,Y)$ as the joint distribution of the source and downstream tasks, respectively. Random variable $X$ corresponds to the input (e.g., materials) and $Y$ is the label (e.g., corresponding values for a specific property). Negative transfer can result from the divergence $d(P_S(X,Y), P_T(X,Y))$, where $d(\cdot, \cdot)$ is a divergence metric over distributions. Wang et al. also argued that the downstream dataset size has a mixed effect on negative transfer; if the downstream task is too small, then it becomes difficult for the learning algorithm to properly learn the similarity between the source and target tasks. Yet, if the downstream task is too large, then transferring from a source task with even a slightly different joint distribution could harm generalization and perform worse than STL.
    
\begin{figure}[h]%
    \centering
    \includegraphics{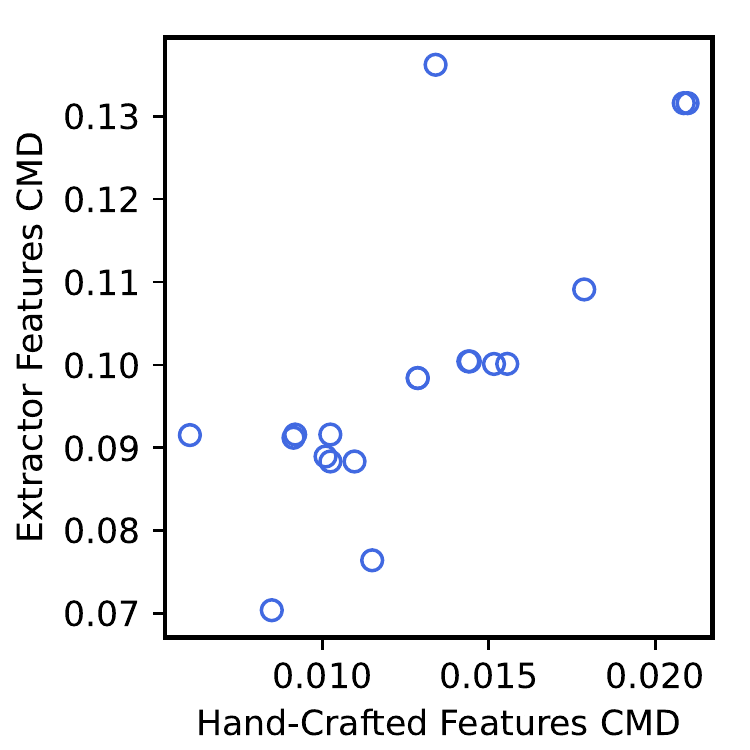}
    \caption{Comparison of CMD in feature space between MP $E_f$ and downstream tasks using a learned featurization versus a procedurally generated featurization. The procedurally generated featurization uses \texttt{matminer.featurizers.site.CrystalNNFingerprint} and \texttt{matminer.featurizers.structure.SiteStatsFingerprint}. 500 structures were subsampled from each dataset during procedural generation for efficiency.}\label{matminer_vs_extractor_figure}
    \end{figure}

    To understand performance variations of MoE across different downstream tasks, we examined the divergence in feature and label space between the source and downstream tasks. In general, atomic structures $X$ and their associated materials properties $Y$ are not independent; the connection between structure and properties is central to materials science. A data-driven comparison of two materials property prediction tasks $S$ and $T$ should thus compare their full joint distributions, $P_S(X,Y)$ and $P_T(X,Y)$. Unfortunately, computing $P_S(X,Y)$, $P_T(X,Y))$, and subsequently $d(P_S(X,Y), P_T(X,Y)))$ is difficult in practice. Instead, we decoupled the feature and label spaces, separately measuring empirical domain and label shifts between source and downstream tasks, $d(P_S(X), P_T(X))$ and $d(P_S(Y), P_T(Y))$. To compute these shifts, we used central moment discrepancy (CMD) \cite{central_moment_discrepancy}, a distance metric for probability distributions on compact space. Intuitively, CMD compares distribution means and arbitrarily high central moments to capture differences in distribution positions and shapes. Up to 50th order central moments were included in our experiments. To measure domain shift, CMD was computed in the learned feature space of the last frozen convolutional layer of each extractor. We found that CMD computed in this feature space showed a strong positive trend with CMD computed in the space of features procedurally generated from local atomic structure order parameters \cite{crystalNN, matminer_site_featurizer}, suggesting the learned feature space faithfully distinguishes atomic structures (see Fig. \ref{matminer_vs_extractor_figure}). To measure label shift, each task's label distribution was first normalized by subtracting the mean and dividing by the standard deviation. Since CMD requires each distribution to be on compact space, a sigmoid function was applied on each dimension of the feature and normalized label spaces to maintain compact support between 0 and 1.
    
\begin{figure}[h]%
    \centering
    \includegraphics{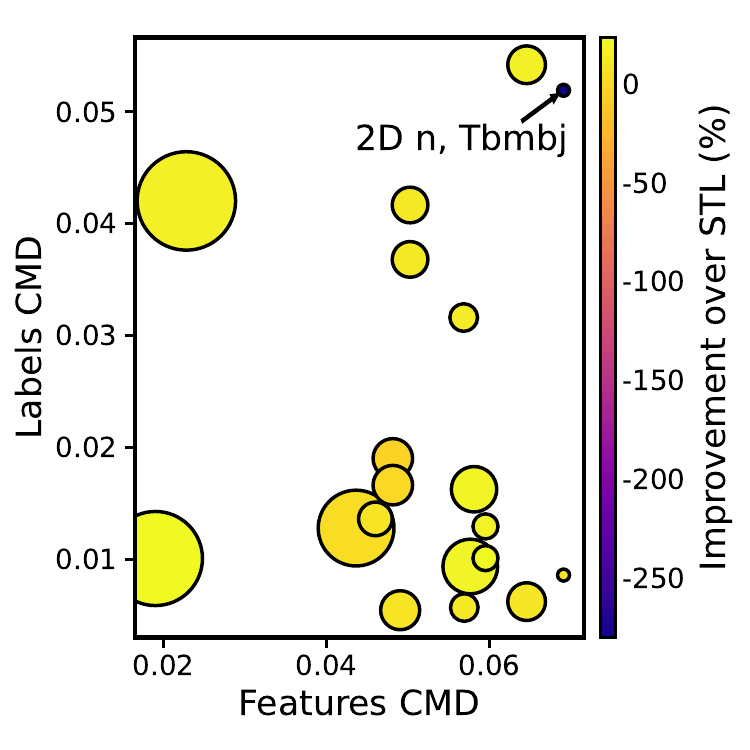}
    \caption{Scatter plot of the average central moment discrepancy (CMD) in label and feature space for each downstream task across the top-$n$ (i.e., ``closest'' $n$) for $n=4$ extractors. Other values of $n$ were also explored without ostensibly different results. Marker sizes are scaled with the downstream dataset size and colored by MoE's improvement on MAE over STL. Another version of this plot without the negative outlier in improvement over STL is found in Table \ref{confetti_party}.}\label{combined_understanding_transfer_plot}
    \end{figure}

    For each downstream task, we plotted the average CMD in label and feature space to the top-$n$ source tasks (i.e., the ``closest'' $n$ source tasks) for $n=4$ (Fig. \ref{combined_understanding_transfer_plot}). Other values of $n$ were explored without yielding ostensibly significant differences. The size of each plotted data point indicates the size of the downstream task data and the color indicates MoE's improvement over STL on that task. While 18 of 19 downstream tasks exhibitted positive transfer under the MoE framework, we observed that the remaining task exhibiting negative transfer had the second highest shift in label space, the highest shift in feature space, and the smallest dataset size across all downstream tasks. This finding suggests that training extractors on diverse materials and properties is a possible avenue for protecting against negative transfer.
    
\begin{figure}[h]%
    \centering
    \includegraphics{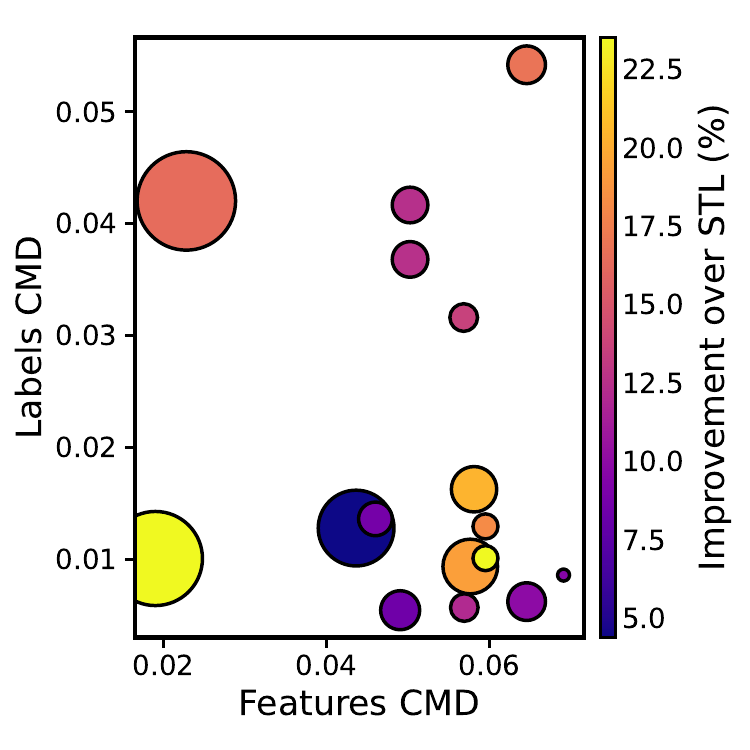}
    \caption{Scatter plot of the average central moment discrepancy (CMD) in label and feature space for each downstream task across the top-4 (i.e., ``closest'' four) extractors. No discernable pattern is evident. Other values of $n$ for top-$n$ were also explored without ostensibly different results. The uncertainty in improvement (not shown) for each data point also obfuscates any trends.}\label{confetti_party}
    \end{figure}

    When the outlier in improvement over STL is removed from Fig. \ref{combined_understanding_transfer_plot}, no discernable trends were observed (see Fig. \ref{confetti_party}). Plausible explanations for the lack of emergent patterns are (1) uncertainty in the improvement over STL obfuscates any trends, (2) 19 downstream tasks is too small of a sample size, (3) CMD is not the optimal distance metric for capturing divergences in task distributions, and (4) decoupling the domain and feature spaces $X$ and $Y$ catastrophically ignores the connection between structure and properties. These obstacles highlight the difficulty in predicting transferability with simple proxies, and that machine learned models, like MoE, are needed to determine which source tasks are most useful for a downstream task.
    
    % \begin{figure}[h]%
    % \centering
    % \includegraphics[width=1.0\textwidth]{features_avg_cmd.png}
    % \caption{\textbf{THIS PLOT IS A PLACEHOLDER. SHOULD WE AXE THIS? Need to break the y-axis, make it not ugly, and stack it on top or next to the same plot for label shift.} Scatter plot of MoE improvement on MAE of MoE over STL versus average central moment discrepancy between features extracted from each downstream task dataset and all source task datasets.}\label{domain_shift_plot}
    % \end{figure}
    
    % \begin{figure}[h]%
    % \centering
    % \includegraphics[width=1.0\textwidth]{labels_avg_cmd.png}
    % \caption{\textbf{THIS PLOT IS A PLACEHOLDER. SHOULD WE AXE THIS? Need to break the y-axis, make it not ugly, and stack it on top or next to the same plot for label shift.} Scatter plot of MoE improvement on MAE of MoE over STL versus average central moment discrepancy between labels from each downstream task dataset and all source task datasets.}\label{label_shift_plot}
    % \end{figure}
    
    \subsubsection{Model interpretability}
\begin{figure}[h]%
    \centering
    \includegraphics[width=1.0\textwidth]{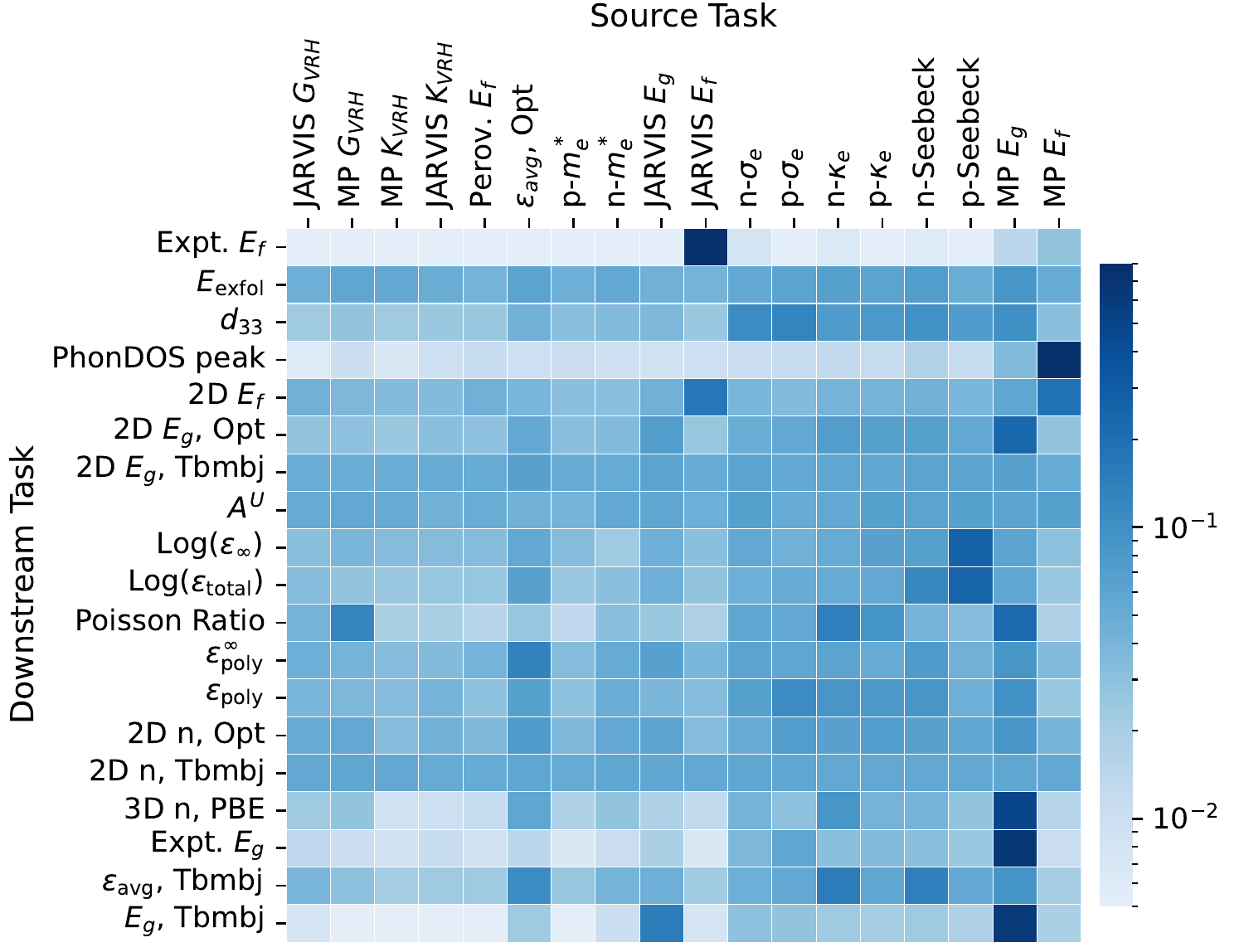}
    \caption{Heatmap of learned probability scores assigned to each feature extractor by our mixture of experts framework. The source tasks are ordered left to right from smallest to largest dataset sizes. Contrary to the heuristic of transferring from the largest source task, our MoE framework did not usually assign the largest score to the largest source task. Instead, the model often assigned scores which were physically intuitive.}\label{pseudoattn_heatmap}
    \end{figure}

    A natural consequence of our MoE gating function (Eq. \eqref{eqG}) is that for each downstream task, the model automatically learns to associate a probability score to each pre-trained extractor. By analyzing these scores, we can readily interpret which pre-trained extractors and source tasks were most relevant to learning each downstream task. In Fig. \ref{pseudoattn_heatmap}, we visualize a heatmap of these learned scores for all downstream tasks.
    
    Despite initializing the model with uniform probability scores assigned to each extractor, we make two notable observations. (1) For a given downstream task, learned scores are relatively robust across different random seeds. (2) The model often learns scores which are physically intuitive rather than simply assigning large scores to extractors pre-trained with more data. For example, when predicting experimental formation energies, the model heavily concentrates probability mass onto the JARVIS and MP formation energy extractors. Dielectric constants computed with the OptB88vDW DFT functional was assigned the highest score for predicting the electronic component of MP dielectric tensors. MP band gaps were assigned the highest score when predicting 2D materials' band gaps and experimental band gaps. Such correspondences suggest our MoE framework's strong performance results in part from learning physically meaningful relationships between source and downstream tasks.
    
    However, there were some instances of counterintuitive task relationships being emitted by the MoE model. For example, while the JARVIS and MP shear moduli extractors were assigned large scores for predicting Poisson ratios as expected, the JARVIS and MP bulk modulus extractors were not. Similarly, when predicting piezoelectric modulus, the MoE model automatically assigned the highest scores to electronic properties like n- and p-type electronic conductivities, electronic thermal conductivities, and Seebeck coefficients, as well as MP band gaps (perhaps because a non-zero band gap is required to maintain electronic polarizations). However, the model did not assign large scores to any mechanical properties. These unexpected results are perhaps better explained by divergences in the datasets' joint distributions in domain and label space rather than by domain knowledge.
    
    \subsubsection{Scaling to an arbitrary number of extractors}
    We anticipate that the number of large materials task datasets, and consequently, the number of potential source tasks, will increase as growing compute resources, new DFT functionals, high-throughput experimental methods, and novel pre-training tasks emerge from the community. To handle many source tasks, we experimented with sparse gating by allowing the sparsity hyperparameter, $k$, from Eq. \eqref{eqTopK} to be less than the number of extractors. During inference, only $k$ extractors would be activated, and thus gradients would only be computed for $k$ extractors in each training iteration. Utilizing sparsity consequently decouples the speed per training iteration from the number of extractors, enabling the MoE framework to scale to an arbitrary number of extractors without concomitant increases in compute cost. In anticipation of the community leveraging performance boosts from model scale \cite{large_chemical_models}, we note that extractors can be distributed across multiple GPUs.
    
    We compared $k$ values of 2, 4, 6, and 10 for three downstream tasks: predicting piezoelectric modulus, 2D exfoliation energies, and experimental formation energies. Surprisingly, we observed no detriment to performance compared to $k=18$, even when setting $k$ as small as 2 (Table \ref{sparse_gating}). This result suggests practitioners can supply our MoE framework with as many pre-trained extractors as desired without fear of increasing compute cost or harming predictive performance.
    
    \begin{table}[h]
    \begin{center}
    \begin{minipage}{193pt}
    \caption{Average test MAE over 5 random seeds for MoE with different settings of the sparsity hyperparameter $k$.}\label{sparse_gating}
    \begin{tabular}{@{}cccc@{}}
    \toprule%
    $k$ & $d_{33}$\footnotemark[1] & $E_\mathrm{exfol}$\footnotemark[2] & Expt. $E_f$\footnotemark[3]\\%
    \midrule
    2   & $0.223\pm0.019$ & $54.4\pm10.7$  & $0.0897\pm0.0124$\\
    4   & $0.216\pm0.021$ & $56.5\pm10.1$  & $0.0874\pm0.0080$\\
    6   & $0.214\pm0.034$ & $55.0\pm12.8$  & $0.0895\pm0.0099$\\
    10  & $0.214\pm0.033$ & $57.0\pm9.7$   & $0.0864\pm0.0093$\\
    18  & $0.208\pm0.029$ & $53.6\pm10.5$  & $0.0908\pm0.0142$\\
    \botrule
    \end{tabular}
    \footnotetext[1]{Piezoelectric modulus from \texttt{Matminer}'s \texttt{piezoelectric\_tensor} dataset \cite{piezo2015}.}
    \footnotetext[2]{Monolayer exfoliation energies (meV/atom) from \texttt{Matminer}'s \texttt{jarvis\_dft\_2d} dataset \cite{jarvis_2d_2017, jarvis2020}.}
    \footnotetext[3]{Experimental formation enthalpies (eV/atom) from \texttt{Matminer}'s \texttt{expt\_formation\_enthalpy} \cite{kim_expt_ef2017} and \texttt{expt\_formation\_enthalpy\_kingsbury} \cite{expt_formation_energy_kingsbury} datasets. The former was preferred when duplicates arose.}
    \end{minipage}
    \end{center}
    \end{table}
    
% \section{Conclusion}
    In conclusion, we presented a mixture of experts framework combining complementary materials datasets and ML models to achieve consistent state-of-the-art performance on a suite of data scarce property prediction tasks. We demonstrated the interpretability of our framework, which readily emits automatically learned relationships between a downstream task and all source tasks in a single training run. We often found these relationships to be physically intuitive. By introducing a sparsity hyperparameter, we also showed that MoE is scalable to an arbitrary number of source tasks and extractors without performance detriment. The MoE framework is general, allowing any model architecture or hand-crafted featurizer to act as extractors and any dataset to act as a source task. We invite the community to engineer new source tasks to train generalizable extractors; explore mixtures of different extractor model classes such as hand-crafted descriptors or equivariant neural networks which predict non-scalar properties; and share materials datasets spanning diverse properties, dataset sizes, and fidelities.
    
\section{Methods}
    \subsection{Crystal graph convolutional neural networks}
    For a full treatment of CGCNNs, see Ref. \cite{cgcnn2018}. Briefly, CGCNNs operate on graph representations of crystals. A crystal structure with $N$ atoms is represented as a graph $G=(\{\boldsymbol{v}_i^0\}_{i=1}^N, \{\boldsymbol{u}_{(i,j)}\}_{i,j=1}^N)$ with initial node features $\boldsymbol{v}_i^0$ representing atom $i$ and edge features $\boldsymbol{u}_{(i,j)}$ representing bond(s) features between atoms $i$ and $j$. In the original implementation, $\boldsymbol{v}_i^0$ is a trainable linear transformation of vectorized elemental features like group number and electronegativity.
    
    Node/atom features are sequentially updated with graph convolutional layers, passing information from node features and shared edges of locally neighboring atoms. CGCNN implements their graph convolutional layer as
    \begin{align}
        \boldsymbol{v}_i^{(t+1)} &= \boldsymbol{v}_i^{(t)} + \sum_{j,k} \sigma(\boldsymbol{z}^{(t)}_{(i,j)_k} \boldsymbol{W}_f^{(t)} + \boldsymbol{b}_f^{(t)}) \odot g(\boldsymbol{z}^{(t)}_{(i,j)_k} \boldsymbol{W}_s^{(t)} + \boldsymbol{b}_s^{(t)})\\
        \boldsymbol{z}^{(t)}_{(i,j)_k} &= \boldsymbol{v}_i^{(t)} \oplus \boldsymbol{v}_j^{(t)} \oplus \boldsymbol{u}_{(i,j)_k}
    \end{align}
    
    where $\boldsymbol{v}_i^{(t)}$ is the node feature of atom $i$ after $t$ graph convolutions, $\sigma(\cdot)$ is a sigmoid function, $g(\cdot)$ is a softplus, $k$ represents the $k$-th bond between atoms $i$ and $j$, $\odot$ is elementwise multiplication, $\oplus$ is concatenation, and $\boldsymbol{W}_f^{(t)}$,  $\boldsymbol{W}_s^{(t)}$, $\boldsymbol{b}_f^{(t)}$, and $\boldsymbol{b}_s^{(t)}$ are trainable parameters for the $t$-th graph convolutional layer. After $T$ convolutional layers, all node features are averaged to produce a feature vector $\boldsymbol{v}_c$ representing the entire crystal. Finally, $\boldsymbol{v}_c$ is passed as input to a multilayer perceptron to yield a prediction.
    
    \subsection{Model training}
    Datasets were split into 70\% training, 15\% validation, and 15\% testing data for 5 random seeds. For each dataset, the same 5 random splits were re-used across STL, TL, and MoE experiments for consistency. Models were trained for 1000 epochs unless validation error did not improve for 500 epochs, in which case early stopping was applied. Models were optimized with Adam, mean squared error loss, a Cosine Annealing scheduler, and a batch size of 250 (or the entire training split - whichever was smaller) on NVIDIA Tesla V100 and A100 GPUs. Each dataset's regression labels were normalized by subtracting the mean and dividing by the standard deviation of labels in the training and validation sets.
    
    During STL and extractor training, all layers were updated with an initial learning rate of 1e-2. During TL and MoE training, all extractor layers were frozen except for the last convolutional layer, which was updated with an initial learning rate of 5e-3. Head layers were updated with an initial learning rate of 1e-2.
    
    Batches were always sampled with uniform random sampling, except when pre-training extractors on the n- and p-type electronic thermal and electronic conductivity source tasks, which had heavily skewed label distributions. For those tasks, batches were sampled with weighted sampling. Specifically, label distributions were split into 30 bins, and the sampling weight for bin $i$ was computed as
    \begin{align*}
    \frac{1/(\mathrm{number\ of\ examples\ in\ bin\ } i)}{\sum\limits_{j \in I_b} 1/(\mathrm{number\ of\ examples\ in\ bin\ } j)}
    \end{align*}
    where $I_b$ represents the set of bin indices with at least one example. Bins with no examples were reassigned a sampling weight of 0.
    
\begin{figure}[h]%
    \centering
    \includegraphics[width=\textwidth]{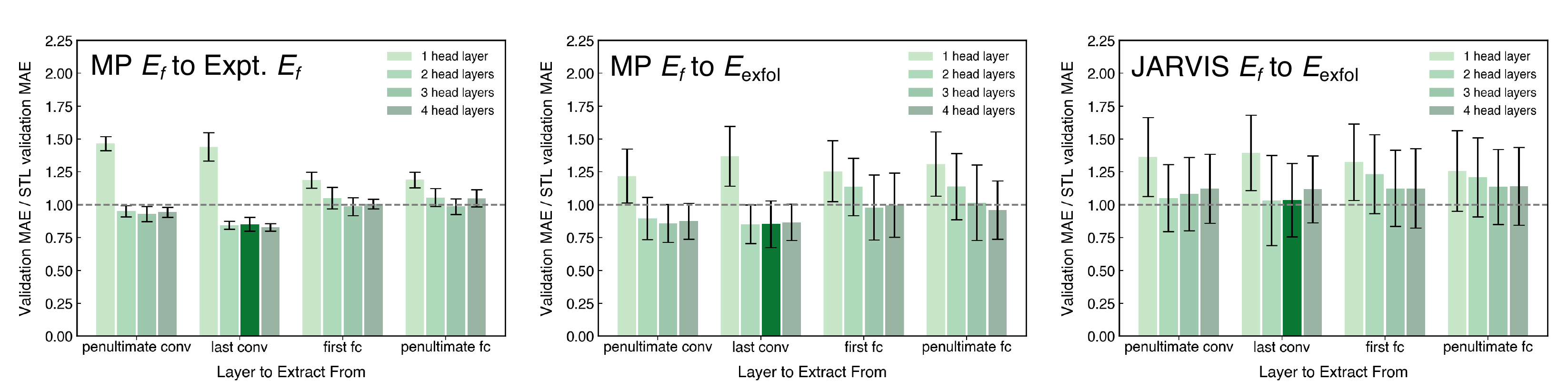}
    \caption{Bar plots of pairwise transfer learning validation error (normalized by STL validation error on the same random split) with various source and downstream tasks when extracting from different frozen, pre-trained layers and when applying different head sizes. Penultimate conv, last conv, first fc, and penultimate fc denote the second to last convolutional layer, the last convolutional layer, the first head layer, and the second to last head layer, respectively. The darkest color bar indicates the chosen hyperparameter setting. 3 head layers were preferred to match the original CGCNN architecture.}\label{layer_to_extract_from}
    \end{figure}

\begin{figure}[h]%
    \centering
    \includegraphics[width=\textwidth]{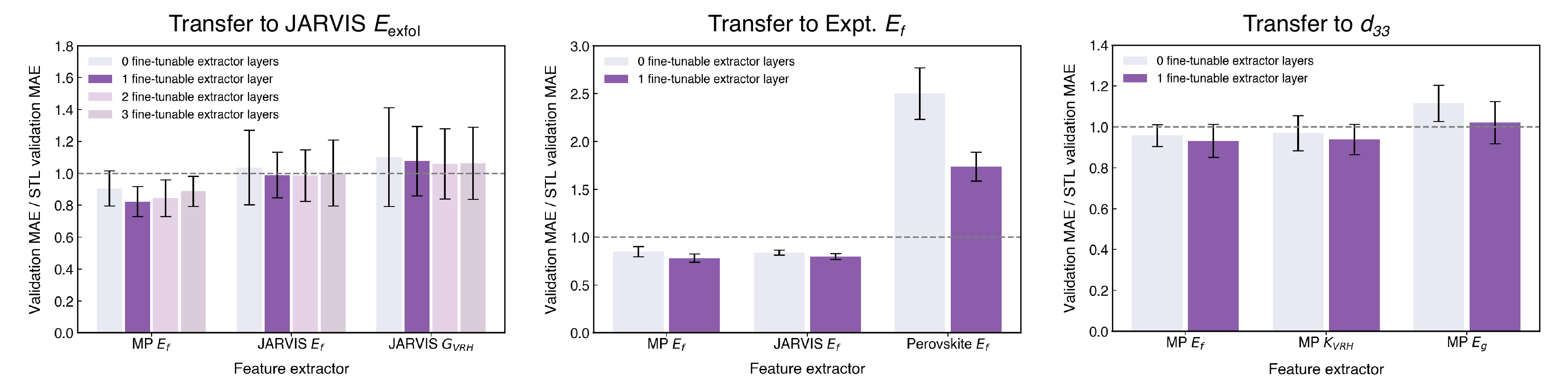}
    \caption{Bar plots of pairwise transfer learning validation error (normalized by STL validation error on the same random split) when fine-tuning different numbers of pre-trained extractor layers with various source and downstream tasks. The darkest color bar indicates the chosen hyperparameter setting.}\label{n_layers_to_finetune}
    \end{figure}

    While some hyperparameter tuning was conducted for pairwise TL (see Figures \ref{layer_to_extract_from} and \ref{n_layers_to_finetune}), we did not do any hyperparameter tuning for MoE, instead drawing the same hyperparameters from TL or from literature. Thus the strong performance of MoE is likely robust and can possibly be further improved with hyperparameter tuning.

\section{Data and Code Availability}
    All data is available from \texttt{Matminer} \cite{matminer2018}. Downstream task datasets, training, validation, and test splits, and code are available at \url{https://github.com/rees-c/MoE}.

%%=============================================%%
%% For presentation purpose, we have included  %%
%% \bigskip command. please ignore this.       %%
%%=============================================%%

\backmatter

\bmhead{Acknowledgments}
This material is based upon work supported by the National Science Foundation under Grant No. 1922758 and utilizes computational resources supported by the National Science Foundation’s Major Research Instrumentation program, grant \#1725729, as well as the University of Illinois at Urbana-Champaign.

\section*{Competing Interests}
The authors declare no competing interests.

\section*{Author contributions}
 R.C., Y.-X.W., and E.E. conceived the idea. R.C. wrote the code, carried out the experiments, and performed the analysis. Y.-X.W. and E.E. supervised and guided the project. The manuscript was prepared by R.C. All authors reviewed and edited the manuscript.


\bibliography{sn-bibliography}% common bib file
% \input{sn-article.bbl}


\section*{Supplementary information}
    \begin{table}[h]
    \begin{center}
    \begin{minipage}{315pt}
    \caption{Source tasks are reported along with test mean absolute errors (MAEs) and their mean absolute deviation (MAD)-normalized values during pre-training.}\label{extractor_tasks}%
    \begin{tabular}{@{}llll@{}}
    \toprule
    Source task & Dataset size & Test MAE & Test MAE/MAD\\
    \midrule
    MP $E_f$\footnotemark[1]              & 132,752                   & 0.0352    & 0.0350\\
    MP $E_g$\footnotemark[2]              & 106,113                    & 0.243     & 0.183\\
    MP $G_{VRH}$\footnotemark[3]          & 10,987                    & 0.0931    & 0.317\\
    MP $K_{VRH}$\footnotemark[4]          & 10,987                    & 0.0749    & 0.259\\
    n-type $\sigma_e$\footnotemark[5]    & 37,390                    & 1.05     & 0.106\\
    p-type $\sigma_e$\footnotemark[6]   & 37,390                    & 0.989     & 0.0605\\
    n-type $\kappa_e$\footnotemark[7]   & 37,390                    & 0.545     & 0.129\\
    p-type $\kappa_e$\footnotemark[8]   & 37,390                    & 0.611     & 0.125\\
    p-type $S$\footnotemark[9]          & 37,390                    & 105.      & 0.325\\
    n-type $S$\footnotemark[10]          & 37,390                    & 99.0      & 0.389\\
    n-type $\overline{m}^*_e$\footnotemark[11]  & 21,037             & 0.552     & 0.610\\
    p-type $\overline{m}^*_h$\footnotemark[12]   & 20,270             & 0.706     & 0.711\\
    Perovskite $E_f$\footnotemark[13]       & 18,928                    & 0.0488    & 0.0863\\
    JARVIS $E_f$\footnotemark[14]           & 25,923                    & 0.0563    & 0.0650\\
    JARVIS dielectric constant (Opt)\footnotemark[15]     & 19,027    & 0.285     & 0.245\\
    JARVIS $E_g$\footnotemark[16]           & 23,455                    & 0.277     & 0.224\\
    JARVIS $G_{VRH}$\footnotemark[17]       & 10,855                    & 0.238     & 0.354\\
    JARVIS $K_{VRH}$\footnotemark[18]      & 11,028                    & 0.105     & 0.160\\
    \botrule
    \end{tabular}
    \footnotetext{All datasets were acquired through \texttt{Matminer} \cite{matminer2018}. Examples with NaN were discarded.}
    \footnotetext[1]{Formation energies from Materials Project [eV/at] \cite{materialsproject2013}}
    \footnotetext[2]{PBE band gaps from Materials Project [eV] \cite{materialsproject2013}}
    \footnotetext[3]{VRH-average shear modulus from Materials Project [log$_{10}$(GPa)] \cite{materialsproject2013}}
    \footnotetext[4]{VRH-average bulk modulus from Materials Project [log$_{10}$(GPa)] \cite{materialsproject2013}}
    \footnotetext[5]{Average conductivity eigenvalue with $10^{-18}$ electrons/cm$^{-3}$ at 300K [log(1+(1/$\Omega$/m/s))] \cite{ricci2017}}
    \footnotetext[6]{Average conductivity eigenvalue with $10^{-18}$ holes/cm$^{-3}$ at 300K [log(1+(1/$\Omega$/m/s))] \cite{ricci2017}}
    \footnotetext[7]{Average eigenvalue of electrical thermal conductivity with $10^{-18}$ carriers/cm$^{-3}$ (n-type) at 300K [log(1+(W/K/m/s))] \cite{ricci2017}}
    \footnotetext[8]{Average eigenvalue of electrical thermal conductivity with $10^{-18}$ carriers/cm$^{-3}$ (p-type) at 300K [log(1+(W/K/m/s))] \cite{ricci2017}}
    \footnotetext[9]{Average eigenvalue of the Seebeck coefficient with $10^{-18}$ holes/cm$^{-3}$ [$\mu$ V/K] \cite{ricci2017}}
    \footnotetext[10]{Average eigenvalue of the Seebeck coefficient with $10^{-18}$ electrons/cm$^{-3}$ [$\mu$ V/K] \cite{ricci2017}}
    \footnotetext[11]{Average eigenvalue of conductivity effective mass with $10^{-18}$ electrons/cm$^{-3}$ (n-type) at 300K [log(1+Unitless)]. Values greater than 100 were discarded. \cite{ricci2017}}
    \footnotetext[12]{Average eigenvalue of conductivity effective mass with $10^{-18}$ holes/cm$^{-3}$ at 300K [log(1+Unitless)]. Values greater than 100 were discarded. \cite{ricci2017}}
    \footnotetext[13]{Heat of formation. The reference state for oxygen was computed from oxygen's chemical potential in water vapor [eV/at] \cite{castelli2012}}
    \footnotetext[14]{Formation energy per atom from the JARVIS database [eV/at] \cite{jarvis2020}}
    \footnotetext[15]{Average of static dielectric functions in x, y, and z directions calculated with the OptB88vDW functional [log(1+Unitless)] \cite{jarvis2020}}
    \footnotetext[16]{Band gaps from the JARVIS database [eV] \cite{jarvis2020}}
    \footnotetext[17]{VRH-average shear moduli from the JARVIS database [log(1+GPa)] \cite{jarvis2020}}
    \footnotetext[18]{VRH-average bulk moduli from the JARVIS database [log(1+GPa)] \cite{jarvis2020}}
    \end{minipage}
    \end{center}
    \end{table}
    
    \begin{table}[h]
    \centering
    \begin{minipage}{100pt}
    \label{model_hyperparameters}%
    \caption{}
    \begin{tabular}{|lc|}
    \hline
     \multicolumn{2}{|c|}{\textbf{Model Hyperparameters}}\\
     \hline
     orig\_atom\_fea\_len  & 92  \\
     \hline
     n\_conv                &  4 \\
     \hline
     nbr\_fea\_len          &  41 \\
     \hline
     atom\_fea\_len         & 64\\
     \hline
     n\_h                   & 1 \\
     \hline
     h\_fea\_len            & 32 \\
     \hline
     \multicolumn{2}{|c|}{\textbf{Data Hyperparameters}}\\
     \hline
     max\_num\_nbr  & 12  \\
     \hline
     radius       & 8   \\
     \hline
     dmin         & 0   \\
     \hline
     step         & 0.2 \\
    \hline
    \end{tabular}
    \end{minipage}
    \end{table}
    
\end{document}